# Low-Temperature Properties of Ising Antiferromagnet on a Stacked Triangular Lattice

M. Žukovič*, L. Mižišin, A. Bobák

Institute of Physics, Faculty of Sciences, P. J. Šafárik University, Park Angelinum 9, 041 54 Košice, Slovakia

An Ising antiferromagnet on a stacked triangular lattice in zero field is studied by Monte Carlo simulations, focusing on the character of the low-temperature phase and the effect of the relative strength of the exchange interaction in the stacking direction α. Our results support the presence of the 3D Wannier phase, with the sublattice magnetization structure ($m$, $-m$, 0) and power-law decaying $m$ with the lattice size. The extent of this low-temperature phase shrinks with decreasing α, however, it appears even at very low values if it is accessed from higher temperatures by sufficiently slow cooling.

## 1. Introduction

An Ising antiferromagnet on a stacked triangular lattice (IASTL) is a relatively simple geometrically frustrated spin model with long history of investigation [1-6]. Nevertheless, its behavior remains an object of controversy even in zero field. There is a broad consensus that at higher temperatures the model exhibits a phase transition from a paramagnetic phase to a partially disordered phase with two sublattices ordered and one disordered of the type ($m$, $-m$, 0). However, as the temperature is lowered, the Landau-Ginzburg-Wilson, Monte Carlo and Monte Carlo mean-field approaches [2,6] predicted transition to a ferrimagnetic state with one sublattice fully ordered and two partially disordered with the structure ($m$, $-m/2$, $-m/2$) and possibility of unsaturated $m < 1$ due to the kinetic effect. On the other hand, some other studies [3-5] argued that Landau-type arguments are unreliable at low temperatures and that the low-temperature phase is a 3D analog of the 2D Wannier phase. Namely, that all the spin chains are fully ordered in the stacking direction, and most configurations (but not all) are such that the chains on two sublattices align antiparallel while those on the third one point in a random direction. Therefore, the character of the low-temperature phase is still not quite clear.

In our study we attempt to shed some more light on the above issues by extensive Monte Carlo simulations. In particular, we focus on the behavior of the sublattice magnetizations in a wide range of the inter- to intra-layer exchange interaction ratio α. We also investigate how α affects the lack of saturation in the zero-temperature sublattice magnetizations.

## 2. Model and simulations

The model of the Ising antiferromagnet on a stacked triangular lattice is described by the Hamiltonian

$$H = -J_1 \sum_{\langle i,j \rangle} s_i s_j - J_2 \sum_{\langle i,k \rangle} s_i s_k, \qquad (1)$$

where $s_i = \pm 1$ is an Ising spin, $\langle i,j \rangle$ and $\langle i,k \rangle$ denote the sums over nearest neighbors in the triangular plane and in adjacent planes, respectively. We choose the exchange interaction parameters $J_1 < 0$ and $J_2 > 0$, so that each of the planes is antiferromagnetic but ferromagnetically coupled to adjacent planes, and define parameter $\alpha = J_2/|J_1|$.

Simulated spin systems are of the size $L^3$, with L ranging from 12 up to 60, and the periodic boundary conditions imposed. The updating follows the Metropolis dynamics and for thermal averaging we typically consider up to $N = 10^6$ Monte Carlo sweeps after discarding another 10% of the sweeps for thermalization. Simulations start from some temperature $t = k_B T/|J_1|$ in the paramagnetic phase and proceed to temperatures gradually decreased by the step $\Delta t$, using the last configuration at $t$ as an initial state at $t-\Delta t$. The triangular lattice can be viewed as consisting of three interpenetrating sublattices A, B and C (spins in the stacking direction belong to the same sublattice). The sublattice magnetizations per spin can be calculated as

$$m_X = 3\langle M_X \rangle / L^3 = 3\langle \sum_{i \in X} s_i \rangle / L^3, \qquad (2)$$

where X = A, B, C and $\langle \ldots \rangle$ denotes thermal averages.

## 3. Results and discussion

In zero field, the IASTL model is known to undergo a phase transition from the paramagnetic phase to a partially disordered one, with the sublattice magnetizations ($m_A$, $m_B$, $m_C$) = ($m$, $-m$, 0) [1-6]. This transition is apparent in the temperature dependencies of the sublattice magnetizations shown in Fig. 1(a), for $\alpha = 1$, $\Delta t = 0.1$ and $L = 30$. At low temperatures, however, the sublattice magnetizations do not saturate to (1, -1, 0). As the temperature is lowered, at some temperature $t_k \approx 0.5$ the magnitudes of $m_A$ and $m_B$ abruptly decrease and seem to ``freeze'' to some nontrivial values which are retained virtually unchanged down to near zero temperature ($t = 0.001$). We found that in this phase, the spins within the chains in the stacking direction align, albeit there is no order among the chains in the *xy* plane on one sublattice and only partial order on the remaining two sublattices. However, by running another simulation using different initialization we obtain generally different saturation values of $m_A$, $m_B$ and $m_C$, for $t \to 0$. Therefore, due to their stochastic character, we performed 100 simulations and statistically evaluated their behavior. In the histograms of the respective values, presented in Fig. 1(b), we can see that $(m_A, m_B, m_C) \approx (0.6, -0.6, 0)$. Hence, this result is apparently in contradiction with the conclusion that the low-temperature phase is ordered with the sublattice magnetizations $(m, -m/2, -m/2)$ [2,6]. It rather favors the partial ordering scenario with the sublattice magnetizations $(m, -m, 0)$ and $m < 1$.

*corresponding author; e-mail: milan.zukovic@upjs.sk

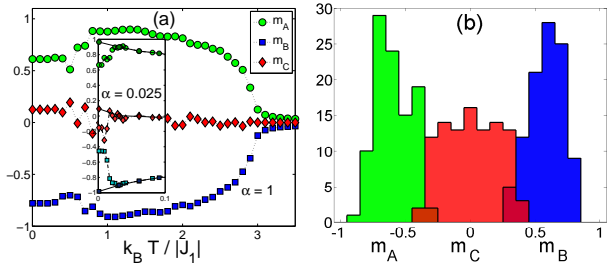

Fig.1. (a) Temperature variations of the sublattice magnetizations for α = 1 with Δ$t$ = 0.1 and L = 30. Inset shows the behavior for α = 0.025 with two different cooling rates of Δ$t$ = 0.03 (solid line) and Δ$t$ = 0.004 (dotted line)  (b) Histograms of the resulting saturation values of $m_A$, $m_B$ and $m_C$ from 100 simulation runs for α = 1.

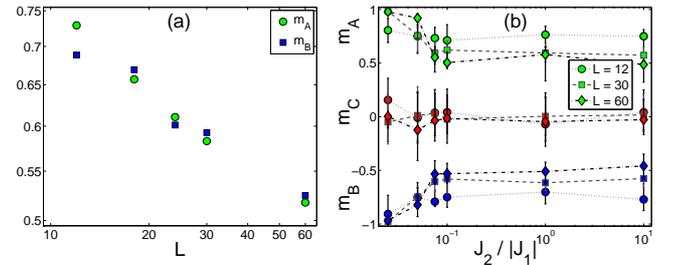

Fig.2. (a) Lattice size dependence of the mean saturation values of $m_A$ and $m_B$ from 100 simulation runs for α = 1, plotted on a log-log scale. (b) Mean saturation values of the sublattice magnetizations as functions of α, for Δ$t$ = 0.1 and different values of L.

The above simulations are performed for a fixed lattice size of L = 30. However, by changing L we observed that the magnitudes of the frozen sublattice magnetizations tend to decrease with increasing lattice size. The plots of the lattice size dependences of $m_A$ and $m_B$ on a log-log scale in Fig. 2(a) indicate the power-law scaling. This however means that in the thermodynamic limit all the sublattice magnetizations vanish and the low-temperature phase shows no long-range ordering. Thus the present result supports the scenario that the low-temperature phase is a 3D analog of the 2D Wannier phase [3-5].

Furthermore, we investigated how the lack of saturation of the zero-temperature sublattice magnetizations is affected by α. In Fig. 2(b) we plot the mean values of the sublattice magnetizations $m_A$, $m_B$ and $m_C$ at $t = 0.001$, reached by the cooling rate Δ$t$ = 0.1, from 100 simulations as functions of the exchange ratio α for different values of L. While for small enough lattice sizes (L = 12) the curves are almost flat, as L increases two distinct regimes of the behavior emerge. Namely, above α ≈ 0.07 the sublattice magnetizations $m_A$ and $m_B$ clearly fail to saturate with the values far from ±1 (close to ±1/2 for L = 60), showing little variation with α. On the other hand, below α ≈ 0.07 the magnitudes sharply increase up to close to ±1. However, if the cooling is performed using a smaller temperature step Δ$t$, the unsaturation effect persists also at these values of α. This is evident from the inset of Fig. 1(a), where the sublattice magnetization behavior is shown for α = 0.025 on approach to zero temperature, using two different cooling rates Δ$t$ = 0.03 and 0.004. While the former leads to the saturation values of $m_A = 1$ and $m_B = -1$, the latter clearly fails to saturate with $m_A \approx 0.65$ and $m_B \approx -0.5$, i.e., the values similar to the case of larger α. Consequently, if we used slower cooling rates, such as Δ$t$ = 0.004, all the curves in Fig. 2(b) would become flat, i.e., almost independent on α.

## 4. Conclusions

In summary, we studied an Ising antiferromagnet on a stacked triangular lattice in zero field by Monte Carlo simulations. We focused on the nature of the low-temperature phase and the lack of saturation in the sublattice magnetizations as zero temperature is approached. Our results support the scenario of the 3D Wannier phase, with the structure $(m, -m, 0)$ and power-law decaying $m$ with the lattice size. This behavior did not seem to be affected by the value of the exchange interaction ratio α, as long as the low-temperature phase is reached by sufficiently slow cooling.

### Acknowledgement

This work was supported by the Scientific Grant Agency of Ministry of Education of Slovak Republic (Grant No. 1/0234/12). The authors acknowledge the financial support by the ERDF EU (European Union European Regional Development Fund) grant provided under the contract No. ITMS26220120047 (activity 3.2.).